\documentclass[oribibl]{llncs}
\usepackage{multicol}
\usepackage{amsmath}
\usepackage{amssymb}
\usepackage{graphicx}
\usepackage{color}
\usepackage{hhline}
\usepackage{times}

\newcommand{\ignore}[1]{}

\newcommand{\mf}[1]{\mathfrak{ #1}}

\newcommand{\bblue}[1]{#1}
\newcommand{\re}[1]{#1}
\newcommand{\bl}[1]{#1}
\newcommand{\gr}[1]{#1}

\def\Li{\mathrm{\Phi}}

\newcommand{\boxtheorem}{  \hfill $\Box$}
\newcommand{\nit}[1]{{\it #1}}

\newcommand{\mc}[1]{\mathcal{ #1}}

\newcommand{\phill}{\phantom{po}  \hfill}

\newcommand{\shp}{\#{\it P}}

\newcommand{\bcq}{BCQ}
\newcommand{\cq}{CQ}

\newcommand{\fo}{FO}

\title{From Database Repairs to Causality in Databases and Beyond \vspace{-4mm}}

\author{{\bf Leopoldo Bertossi\vspace{-2mm}}\thanks{Email: leopoldo.bertossi@skema.edu. \ Member of the Millennium Inst. for Foundational Research on Data (IMFD, Chile)}}
\institute{\bf SKEMA Business School, Montreal, Canada\vspace{-3mm}}

\begin{document}

\thispagestyle{empty}
\pagestyle{plain}
\maketitle

\begin{abstract}
We describe some recent approaches to score-based explanations for query answers in databases\ignore{ and outcomes from classification models in machine learning}. The focus is on work done by the author and collaborators.  Special emphasis is placed on \ignore{declarative approaches based on answer-set programming to} the use of counterfactual reasoning for score specification and computation. Several examples that illustrate the flexibility of these methods are shown. \vspace{-2mm}
\end{abstract}

\section{Introduction}

\vspace{-1mm}
In data management one wants {\em explanations} for certain results. For example, for query results from databases. Explanations, that may come in different forms, have been the subject of philosophical enquires for a long time, but, closer to our discipline, they appear under different forms in model-based diagnosis and in causality as developed in artificial intelligence.

In the last few years, explanations that are based on {\em numerical scores} assigned to elements of a model that may contribute to an outcome have become popular. These scores attempt to capture the degree of contribution of those components to an outcome, e.g. answering questions like these: What is the contribution of this tuple to the answer to this query?

Different scores have been proposed in the literature, and some that have a relatively older history have been applied. Among the latter we find the general {\em responsibility score} as found in {\em actual causality} \cite{HP05,CH04}. For a particular kind of application, one has to define the right  causality setting, and then apply the responsibility measure to the participating variables (see \cite{halpern15} for an updated treatment of causal responsibility).

In data management,  responsibility has been used to
quantify the strength of a tuple  as a cause for a query result \cite{suciu,tocs} (see Section \ref{sec:causal}). \ The {\em responsibility score}, $\nit{Resp}$,  is
based on the notions of {\em counterfactual intervention} as appearing in actual causality. More specifically,
(potential) executions of   {\em counterfactual interventions} on a {\em structural logico-probabilistic model} \ \cite{HP05} are investigated, with the purpose of answering hypothetical  questions of the form: \ {\em What would happen if we change ...?}.

Database repairs are commonly used to define and obtain semantically correct query answers from a database that may fail to satisfy a given set of integrity constraints (ICs) \cite{bertossiSynth}. A connection between repairs and actual causality in DBs has been used to obtain complexity results and algorithms for  the latter \cite{tocs} (see Section \ref{sec:ICs}).

The {\em Causal Effect} score is also based on causality, mainly for  {\em observational studies} \cite{rubin,holland,pearl}. It has been applied in data management in \cite{tapp16} (see Section \ref{sec:CE}).

The Shapley value, as found in {\em coalition game theory} \cite{S53}, has been used for the same purpose \cite{LBKS20,SigRec21}. Defining the right game function, the {\em Shapley value} assigned to a player reflects its contribution to the wealth function.
The Shapley value, which  is firmly established in game theory, and is also used in several other areas \cite{S53,R88}.\ The main idea is that {\em several tuples together}, much like
{players in a coalition game}, are necessary to  produce a query result. Some may contribute more than others to the {\em  wealth distribution function} (or simply,  game function), which in this case becomes the query result, namely $1$ or $0$ if the query is Boolean, or a number if we have an aggregation query.
  This use of Shapley value was developed in \cite{LBKS20,SigRec21} (see Section \ref{sec:shapy}).

In this article we survey some of the recent advances on the use and computation of the above mentioned  score-based explanations for query answering in databases. This is not intended to be an exhaustive survey of the area. Instead, it is heavily influenced by our latest research. \ignore{ Taking advantage of  repairs, we also show how to specify and compute a numerical measure of inconsistency of database \cite{lpnmr19}. In this case, this would be  a {\em global} score, in contrast with the {\em local} scores applied to individual tuples in a database or feature values in an entity. \ }
 To introduce the concepts and techniques we will use mostly examples, trying  to convey the main intuitions and issues.

This paper is structured as follows. In Section \ref{sec:back}, we provide some preliminaries on databases. In Section \ref{sec:dbs}, we introduce causality in databases and the responsibility score, and also the causal effect score. In Section \ref{sec:repCon}, we show the connection between causality in databases and database repairs. In Section \ref{sec:ICs}, we show how integrate ICs in the causality setting. In Section \ref{sec:shapy}, we show how to use the Shapley value to provide explanation scores to database tuples in relation to a query result.  In Section \ref{sec:last}, we make some general remarks on relevant open problems.

\vspace{-2.5mm}
\section{Background}\label{sec:back}

\vspace{-2.5mm}
  A relational schema $\mc{R}$ contains a domain of constants, $\mc{C}$,  and a set of  predicates of finite arities, $\mc{P}$. \ $\mc{R}$ gives rise to a language $\mf{L}(\mc{R})$ of first-order (FO)  predicate logic with built-in equality, $=$.  Variables are usually denoted with $x, y, z, ...$; and finite sequences thereof with $\bar{x}, ...$; and constants with $a, b, c, ...$, etc. An {\em atom} is of the form $P(t_1, \ldots, t_n)$, with $n$-ary $P \in \mc{P}$   and $t_1, \ldots, t_n$ {\em terms}, i.e. constants,  or variables.
  An atom is {\em ground} (a.k.a. a tuple) if it contains no variables. A database (instance), $D$, for $\mc{R}$ is a finite set of ground atoms; and it serves as an  \ignore{The {\em active domain} of a database $D$, denoted ${\it Adom}(D)$, is the set of constants that appear in atoms of $D$.} interpretation structure for  $\mf{L}(\mc{R})$.

A {\em conjunctive query} (\cq) is a \fo \ formula,  $\mc{Q}(\bar{x})$, of the form \ $\exists  \bar{y}\;(P_1(\bar{x}_1)\wedge \dots \wedge P_m(\bar{x}_m))$,
 with $P_i \in \mc{P}$, and (distinct) free variables $\bar{x} := (\bigcup \bar{x}_i) \smallsetminus \bar{y}$. If $\mc{Q}$ has $n$ (free) variables,  $\bar{c} \in \mc{C}^n$ \ is an {\em answer} to $\mc{Q}$ from $D$ if $D \models \mc{Q}[\bar{c}]$, i.e.  $Q[\bar{c}]$ is true in $D$  when the variables in $\bar{x}$ are componentwise replaced by the values in $\bar{c}$. $\mc{Q}(D)$ denotes the set of answers to $\mc{Q}$ from $D$. $\mc{Q}$ is a {\em Boolean conjunctive query} (\bcq) when $\bar{x}$ is empty; and when {\em true} in $D$,  $\mc{Q}(D) := \{\nit{true}\}$. Otherwise, it is {\em false}, and $\mc{Q}(D) := \emptyset$. We will consider only conjunctive queries or disjunctions thereof.

We consider as integrity constraints (ICs), i.e. sentences of $\mf{L}(\mc{R})$: (a) {\em denial constraints} \ (DCs), i.e.  of the form $\kappa\!:  \neg \exists \bar{x}(P_1(\bar{x}_1)\wedge \dots \wedge P_m(\bar{x}_m))$,
where $P_i \in \mc{P}$, and $\bar{x} = \bigcup \bar{x}_i$; and (b) {\em inclusion dependencies} (INDs), which are  of the form \ $\forall \bar{x} \exists \bar{y}(P_1(\bar{x}) \rightarrow P_2(\bar{x}^\prime,\bar{y}))$, where $P_1, P_2 \in \mc{P}$, and $\bar{x}^\prime \subseteq \bar{x}$.
\  If an instance $D$ does not satisfy the set $\Sigma$ of ICs associated to the schema, we say that $D$ is {\em inconsistent}, denoted with \ $D \not \models \Sigma$.

\vspace{-2mm}

\section{Causal Explanations in Databases}\label{sec:dbs}

\vspace{-2mm}
 In data management we {need to understand and compute}
{\em  why}  certain results are obtained or not, e.g. query answers,  violations of semantic conditions, etc.; and we
expect a  database system to provide {\em explanations}.

\subsection{Causal responsibility}\label{sec:causal}
\vspace{-1mm}

Here, we will consider
{\em causality-based explanations}  \cite{suciu,suciuDEBull}, which we will illustrate by means of an example.

 \begin{example}  \label{ex:uno}  \ Consider the database ${D}$, and the Boolean conjunctive query (BCQ)

\begin{multicols}{2}

\vspace{4mm}\hspace*{1cm}\begin{tabular}{l|c|c|} \hline
$R$  & $A$ & $B$ \\\hline
 & $a$ & ${b}$\\
& $c$ & $d$\\
& ${b}$ & ${b}$\\
 \hhline{~--}
\end{tabular} \hspace*{0.5cm}\begin{tabular}{l|c|c|}\hline
$S$  & $C$  \\\hline
 & $a$ \\
& $c$ \\
& ${b}$ \\ \hhline{~-}
\end{tabular}

\phantom{oo}

\phantom{oo}

\vspace{-6mm}
\noindent $\mc{Q}\!: \ \exists x \exists y ( S(x) \land R(x, y) \land S(y))$, 
for which \ ${D \models \mc{Q}}$ holds, i.e. the query is true in $D$. \ We ask about the  causes for $\mc{Q}$ to be true.

\phantom{oo}
\end{multicols}

\vspace{-4mm}  \
A tuple ${\tau \in D}$ is
{\em counterfactual cause} for  ${\mc{Q}}$ (being true in $D$) if \ ${D\models \mc{Q}}$ \ and \ ${D\smallsetminus \{\tau\}  \not \models \mc{Q}}$.
\ In this example,   {$S(b)$ is a counterfactual cause for $\mc{Q}$}: \ If ${S(b)}$ is removed from ${D}$,
 ${\mc{Q}}$ is no longer true.

Removing a single tuple may not be enough to invalidate the query. Accordingly, a tuple ${\tau \in D}$ is  an {\em actual cause} for  ${\mc{Q}}$
if there  is a {\em contingency set} \ ${\Gamma \subseteq D}$,  such that \ ${\tau}$ \ is a   counterfactual cause for ${\mc{Q}}$ in ${D\smallsetminus \Gamma}$.
\ In this example,  ${R(a,b)}$ is not a counterfactual cause for ${\mc{Q}}$, but it is an actual cause  with contingency set
${\{ R(b,b)\}}$: \ If ${R(b,b)}$ is removed from ${D}$, ${\mc{Q}}$ is still true, but further removing ${R(a,b)}$ makes ${\mc{Q}}$ false.
\boxtheorem \end{example}

Notice that every counterfactual cause is also an actual cause, with empty contingent set.   Actual causes that are not counterfactual causes need company to invalidate a query result.
 \ Now we ask  how strong are tuples as actual causes. \ To answer  this question, we appeal to the {\em responsibility} of an actual cause ${\tau}$ for ${\mc{Q}}$ \cite{suciu}, defined by:

 \vspace{-2mm}
\begin{equation*}
{\nit{Resp}_{\!_D}^{\!\mc{Q}}\!(\tau) \ := \ \frac{1}{|\Gamma| \ + \ 1}},
\end{equation*}
where ${|\Gamma|}$ is the
size of a smallest contingency set, $\Gamma$, for ${\tau}$, \ and  $0$, otherwise.

\vspace{-2mm}
\begin{example} \ (ex. \ref{ex:uno} cont.) \ The {responsibility of ${R(a,b)}$ is \  $\frac{1}{2}$} ${= \frac{1}{1 + 1}}$ \ (its several smallest contingency sets have all size ${1}$). \
  ${R(b,b)}$ and ${S(a)}$ are also actual causes with responsibility  \ ${\frac{1}{2}}$; and
  ${S(b)}$ is actual (counterfactual) cause with responsibility \   $1$ ${= \frac{1}{1 + 0}}$. \boxtheorem
\end{example}

High responsibility tuples provide more interesting explanations. Causes in this case are tuples that come with their responsibilities as  ``scores".
All tuples can be seen as actual causes, but only those with non-zero responsibility score matter. \ Causality and responsibility in databases can be extended to the attribute-value level \cite{tocs,kais}.

 There are connections between database causality and  {\em consistency-based diagnosis} \ and \ {\em abductive diagnosis}, that are two forms of {\em model-based diagnosis} \cite{struss,rw22}. There are also connections with {\em database repairs} \cite{pods99,bertossiSynth}. These connections have led to complexity and algorithmic results for causality and responsibility \cite{tocs,flairsExt} (see Section \ref{sec:repCon}).

\vspace{-2mm}
\subsection{The causal-effect score}\label{sec:CE}

\vspace{-1mm}
Sometimes, as we will see right here below, responsibility does not provide intuitive or expected results, which led to the consideration of an alternative score, the {\em causal-effect score}. We show the issues and the score by means of an example.

\begin{example} \ \label{ex:ce} Consider the database ${E}$ that represents the graph below, and the Boolean  query $\mc{Q}$ that is true in $E$ if there is a path from ${a}$ to ${b}$. Here, $E \models \mc{Q}$. Tuples have global tuple identifiers (tids) in the left-most column, which is not essential, but convenient.

\vspace{-2mm}
\begin{multicols}{3}

 \hspace*{5mm} {\footnotesize \begin{tabular}{l|c|c|} \hline
 {$E$}  &  ${A}$ &  ${B}$ \\\hline
 {$t_1$} & { $a$} &  {$b$}\\
{$t_2$}&  {$a$} &  {$c$}\\
{$t_3$}&  {$c$} &  {$b$}\\
{$t_4$}&  {$a$} &  {$d$}\\
{$t_5$}&  {$d$} &  {$e$}\\
{$t_6$}&  {$e$} &  {$b$}\\ \cline{2-3}
\end{tabular}}

 \hspace{-10mm} \includegraphics[width=3.3cm]{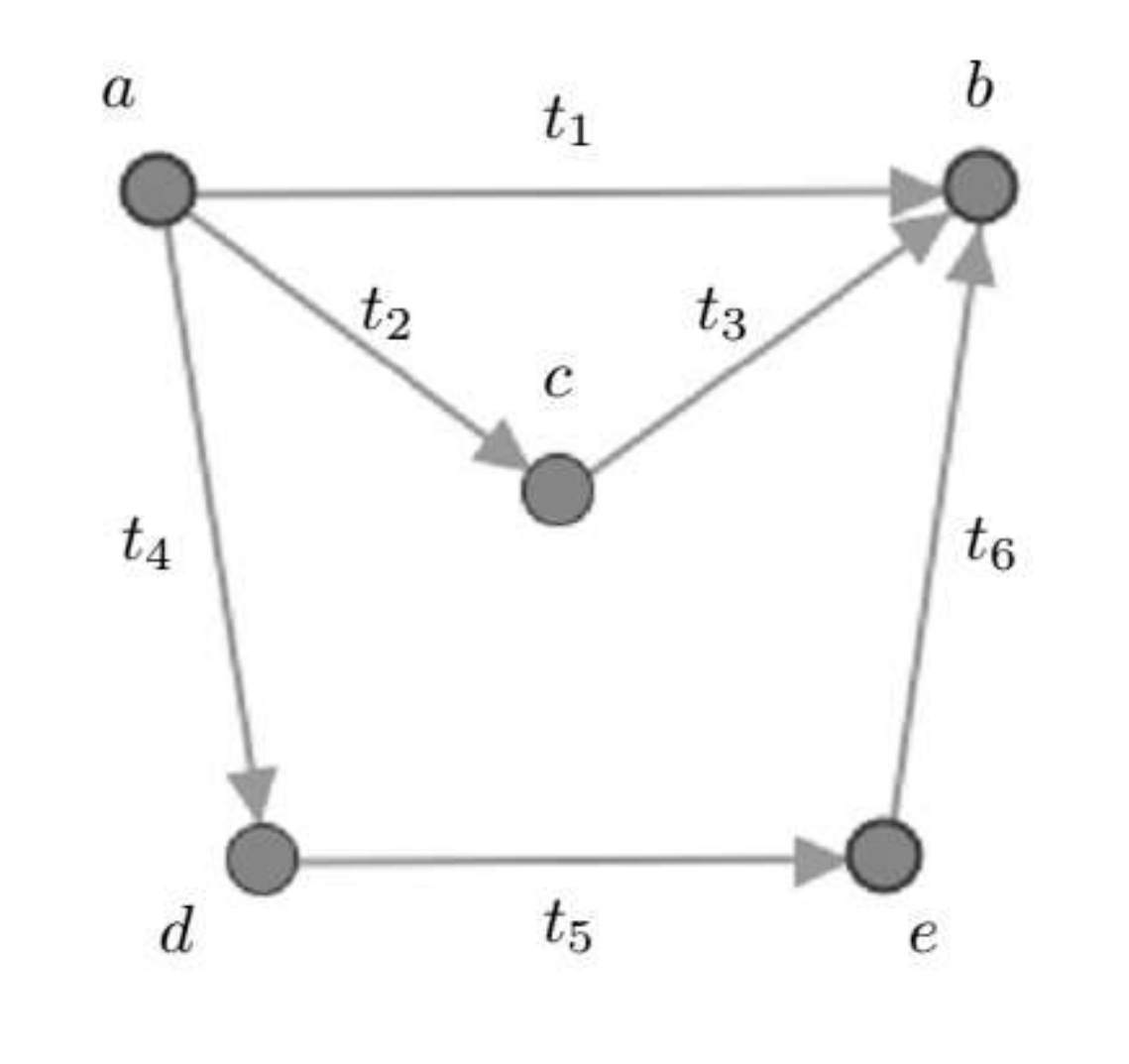}

\ignore{{\small
\begin{eqnarray}
         \nit{yes }& {\leftarrow}&  {P(a,b)} \nonumber \\
        {P(x,y)}& {\leftarrow}& { E(x,y)} \nonumber \\
        {P(x,y)}& {\leftarrow}& { P(x,z), E(z,y)}\nonumber
\end{eqnarray}
}}

{\small
\begin{eqnarray}
         \mc{Q}: \ E(a,b) \  \vee \hspace{2cm} \nonumber\\ \exists x(E(a,x) \wedge E(x,b)) \ \vee \hspace{1.7cm}\nonumber\\
    \hspace*{-1.2cm}   \exists y \exists z(E(a,y) \wedge E(y,z) \wedge E(z,b)) \ \ \ \ \ \ \ \ \ \ \ \ \ \ \ \ \  \nonumber
\end{eqnarray}
}

\end{multicols}

\vspace{-4mm}
All tuples are actual causes since every tuple appears in a path from $a$ to $b$. Also,
all the tuples  have the same causal responsibility, {$\frac{1}{3}$}, which
may be counterintuitive, considering that \ ${t_1}$ provides a direct path from ${a}$ to ${b}$.
\boxtheorem\end{example}

In \cite{tapp16}, the notion  {\em causal effect} was introduced. It is based on three main ideas, namely, the transformation, for auxiliary purposes, of the database into a probabilistic database, the expected value of a query, and
interventions on the lineage of the query \cite{lineage,probDBs}. The lineage of a query represents, by means of a propositional formula, all the ways in which the query can be true in terms of the potential database tuples, and their combinations. Here, ``potential" refers to tuples that can be built with the database predicates and the database (finite) domain. These tuples may belong to the database at hand or not. For a given database, $D$, some of those atoms become true, and others false, which leads to the instantiation of the lineage (formula) o $D$.

 \begin{example} \ Consider the database \ ${D}$  below, and a  BCQ. 

\begin{multicols}{2}
\hspace*{5mm}{\begin{tabular}{c|c|c|}\hline
 $R$ & $A$ & $B$ \\ \hline
  & $a$ & $b$\\
  & $a$ & $c$\\
  & $c$ & $b$\\ \hhline{~--}
  \end{tabular}~~~~~~~~~\begin{tabular}{c|c|}\hline
 $S$ & $C$ \\ \hline
  & ${b}$\\
  & $c$\\
  & \\ \hhline{~-}
  \end{tabular}}

 \noindent ${\mc{Q}: \ \exists x \exists y (R(x, y) \wedge  S(y))}$, which is true in ${D}$.
\end{multicols}

\vspace{-2mm}
For the database $D$ in our example, the lineage of the query { instantiated on ${D}$} is given by the {propositional formula}:
\vspace{-1mm}\begin{equation}
{\Li_\mc{Q}(D)= (X_{R(a, b)} \wedge  X_{S(b)})  \vee (X_{R(a, c)} \wedge  X_{S(c)}) \vee (X_{R(c, b)} \wedge  X_{S(b)})}, \label{eq:lin}
\end{equation}
where   ${X_\tau}$ is a  {propositional variable} that is true iff  ${\tau \in D}$. \ Here,
  ${\Li_\mc{Q}(D)}$ \ takes value ${1}$ in ${D}$.

Now, for illustration, we want to quantify the contribution of tuple ${S(b)}$ to the query answer. \
For this purpose, we assign, uniformly and independently, probabilities to the tuples in ${D}$, obtaining a
 {\em probabilistic database} \ ${D^{{p}}}$ \cite{probDBs}.  \ Potential tuples outside ${D}$ get probability $0$.

\vspace{1mm}
   \hspace*{3cm}
   \begin{tabular}{c|c|c|c|}\hline
 $R^{{p}}$ & $A$ & $B$ & {$\mbox{prob}$}\\ \hline
  & $a$ & $b$ & \scriptsize{${\frac{1}{2}}$}\\
  & $a$ & $c$& \scriptsize{$\frac{1}{2}$}\\
  & $c$ & $b$& \scriptsize{$\frac{1}{2}$}\\ \hhline{~---}
  \end{tabular}~~~~~~~~~\begin{tabular}{c|c|c|}\hline
 $S^p$ & $C$ & $\mbox{prob}$\\ \hline
  & ${b}$& \scriptsize{${\frac{1}{2}}$}\\
  & $c$& \scriptsize{$\frac{1}{2}$}\\
  & & \\ \hhline{~--}
  \end{tabular}

 \vspace{2mm}  {The $X_\tau$'s become independent, identically distributed Boolean random variables}; \ and ${\mc{Q}}$ becomes a Boolean random variable.
Accordingly, we can ask about the probability that $\mc{Q}$ takes the truth value $1$ (or $0$) when an {\em intervention} is performed on $D$.

     Interventions are of the form ${\nit{do}(X = x)}$, meaning making ${X}$ take value ${x}$, with $x \in \{0,1\}$, in the {\em structural model}, in this case, the lineage. That is, we ask,
for \ ${\{y,x\} \subseteq \{0,1\}}$, about the conditional probability  ${P(\mc{Q} = y~|~ {\nit{do}(X_\tau = x)})}$, i.e. conditioned to  making ${X_\tau}$ false or true.

For example, with ${\nit{do}(X_{S(b)} = 0)}$ and $\nit{do}(X_{S(b)} = 1)$, the lineage in (\ref{eq:lin}) becomes, resp., and abusing the notation a bit:

\vspace{-5mm}
\begin{eqnarray*}
\Li_\mc{Q}(D|\nit{do}(X_{S(b)} = 0) &:=&  (X_{R(a, c)} \wedge  X_{S(c)}).\\
\Li_\mc{Q}(D|\nit{do}(X_{S(b)} = 1) &:=& X_{R(a, b)}  \vee (X_{R(a, c)} \wedge  X_{S(c)}) \vee X_{R(c, b)}.
\end{eqnarray*}

\vspace{-2mm}
On the basis of these lineages and  \ ${D^{{p}}}$, \ when ${X_{S(b)}}$ is made false, \ the probability that the instantiated lineage becomes true in {$D^p$} is:

\vspace{1mm}
  \centerline{${P(\mc{Q} = 1~|~ {\nit{do}(X_{S(b)} = 0)}) = P(X_{R(a, c)}=1) \times P(X_{S(c)}=1) = \frac{1}{4}}$.}
\vspace{1mm}
  When ${X_{S(b)}}$ is made true, \ the probability of the lineage  being true in ${D^p}$ is:

\vspace{1mm}
  \centerline{${P(\mc{Q} = 1~|~{\nit{do}( X_{S(b)} = 1)}) = P(X_{R(a, b)}  \vee (X_{R(a, c)} \wedge  X_{S(c)}) \vee X_{R(c, b)} =1)}
{=  \  \frac{13}{16}}.$}

\vspace{1mm}
The {\em causal effect} of a tuple ${\tau}$ is defined by:
\begin{equation*}
{\mathcal{CE}^{D,\mc{Q}}(\tau) \ := \ \mathbb{E}(\mc{Q}~|~\nit{do}(X_\tau = 1)) - \mathbb{E}(\mc{Q}~|~\nit{do}(X_\tau = 0))}.
\end{equation*}

\vspace{-2mm}
In particular, using the probabilities computed so far:

\vspace{-6mm}
\begin{eqnarray*}
\mathbb{E}(\mc{Q}~|~\nit{do}(X_{S(b)} = 0)) &=&  P(\mc{Q} =1~|~\nit{do}(X_{S(b)} = 0)) \ = \  \frac{1}{4},\\
\mathbb{E}(\mc{Q}~|~\nit{do}(X_{S(b)} = 1)) &=&  P(\mc{Q} =1~|~\nit{do}(X_{S(b)} = 1)) \ = \ \frac{13}{16}.
\end{eqnarray*}

\vspace{-2mm}
Then, \  the causal effect for the tuple ${S(b)}$ is:
   ${\mathcal{CE}^{D,\mc{Q}}(S(b)) = \frac{13}{16} - \frac{1}{4} = {\frac{9}{16}} \ > \ 0}$, showing that the tuple is relevant for the query result, with a relevance score provided by the causal effect, of   $\frac{9}{16}$.
\boxtheorem \end{example}

Let us now retake the initial example of this section.

\begin{example} \ (ex. \ref{ex:ce} cont.)  \ The  query has the lineage:
$$\Li_\mc{Q}(D) = X_{t_1} \ \vee \ (X_{t_2}\wedge X_{t_3}) \ \vee \ (X_{t_4} \wedge X_{t_5} \wedge X_{t_6}).$$
It holds: \vspace{-5mm}
\begin{eqnarray*}
\mathcal{CE}^{D,\mc{Q}}(t_1)  &=&  {0.65625},\\
\mathcal{CE}^{D,\mc{Q}}(t_2)  &=&  \mathcal{CE}^{D,\mc{Q}}(t_3) =  0.21875,\\
\mathcal{CE}^{D,\mc{Q}}(t_4)  &=&  \mathcal{CE}^{D,\mc{Q}}(t_5) = \mathcal{CE}^{D,\mc{Q}}(t_6) = 0.09375.
\end{eqnarray*}

\vspace{-3mm}
The causal effects are different for different tuples, and the scores are much more
intuitive than the responsibility scores.  \boxtheorem \end{example}

\vspace{-5mm}
\section{The Database Repair Connection}\label{sec:repCon}

\begin{figure}[h]
\begin{center}
\includegraphics[width=7cm]{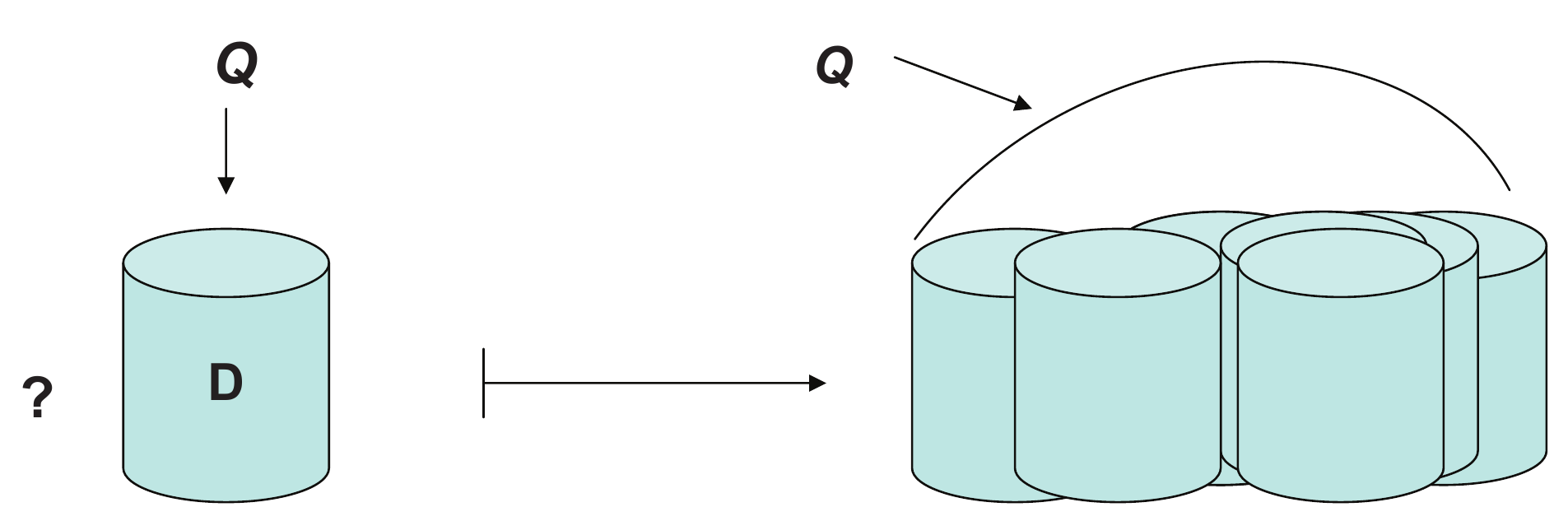}

\centerline{\hspace*{-5mm}$ \re{\not} \models \ \mbox{\bf ICs}$ \hspace{3.5cm} $\models \ \mbox{\bf ICs}$}

\vspace{-2mm}\caption{Database repairs and consistent query answers}\label{fig:reps}
\end{center}
\end{figure}

\vspace{-2mm}
In this section we will first establish a useful connection between database repairs and causes as tuples in a database \cite{pods99,bertossiSynth}.
The notion of {\em repair} of a relational database was introduced in order to formalize the notion of {\em consistent query answering} (CQA), as shown in Figure \ref{fig:reps}: If a database $D$ is inconsistent in the sense that is does not satisfy a given set of integrity constraints, $\nit{ICs}$, and a query $\mc{Q}$ is posed to $D$  (left-hand side of Figure \ref{fig:reps}), what are the meaningful, or consistent, answers to $\mc{Q}$ from $D$? They are sanctioned as those that hold (are returned as answers) from {\em all} the {\em repairs} of $D$. The repairs of $D$ are consistent instances $D'$ (over the same schema of $D$), i.e. $D' \models \nit{ICs}$, and {\em minimally depart} from $D$ (right-hand side of Figure \ref{fig:reps}).

Notice that: (a) We have now a {\em possible-world} semantics for (consistent) query answering; and (b) we may use in principle any reasonable notion of distance between database instances, with each choice defining a particular {\em repair semantics}. In the rest of this section we will illustrate two classes of repairs, which have been used and investigated the most in the literature. Actually, repairs in general have got a life of their own, beyond consistent query answering.

\vspace{-6mm}

\vspace{-11mm}
\begin{example} \label{ex:theEx} Let us consider the following set of {\em denial constraints} (DCs) and a database $D$, whose relations (tables) are shown right here below. $D$ is inconsistent, because it violates the DCs:  it satisfies the joins that are prohibited by the DCs.

\vspace{-6mm}
\begin{multicols}{2}
\begin{eqnarray*}
\neg \exists x \exists y(P(x) \wedge Q(x,y))\\
\neg \exists x \exists y(P(x) \wedge R(x,y))
\end{eqnarray*}

\phantom{ooo}

\begin{tabular}{c|c|}\hline
$P$&A\\ \hline
&a\\
&e\\ \hhline{~-}
\end{tabular}~~~~~~~~
\begin{tabular}{c|c|c|}\hline
$Q$&A&B\\ \hline
& a & b\\ \hhline{~--}
\end{tabular}~~~~~~~~
\begin{tabular}{c|c|c|}\hline
$R$&A&C\\ \hline
& a & c\\ \hhline{~--}
\end{tabular}
\end{multicols}

We want to repair the original instance by {\em deleting tuples} from relations. Notice that, for DCs, insertions of new tuple will not restore consistency. We could change (update) attribute values though, a possibility that has been investigated in \cite{kais}.\ignore{will consider in Section \ref{sec:causAttr}.}

Here we have two {\em subset repairs}, a.k.a. {\em S-repairs}. They are subset-maximal consistent subinstances of $D$: \ $D_1 = \{P(e), Q(a,b), R(a,c)\}$ \ and \ $D_2 = \{P(e),$ $ P(a)\}$. They are consistent, subinstances of $D$, and any proper superset of them (still contained in $D$) is inconsistent. (In general, we will represent database relations as set of tuples.)

We also have {\em cardinality repairs}, a.k.a.  {\em C-repairs}. They are consistent subinstances of $D$ that minimize the {\em number} of tuples by which they differ from $D$. That is, they are maximum-cardinality consistent  subinstances. In this case, only
$D_1$ is a C-repair. \ Every C-repair is an S-repair, but not necessarily the other way around. \boxtheorem
\end{example}

Let us now consider a BCQ
\begin{equation}
\mc{Q}\!: \exists \bar{x}(P_1(\bar{x}_1) \wedge \cdots \wedge P_m(\bar{x}_m)),
\end{equation}
which we assume is  true in  a database $D$. \ It turns out that we can obtain the causes for $\mc{Q}$ \ to be true $D$, and their contingency sets from database repairs. In order to do this, notice that
$\neg \mc{Q}$ \ becomes a  DC \begin{equation}
\kappa(\mc{Q})\!: \   \neg \exists \bar{x}(P_1(\bar{x}_1) \wedge \cdots \wedge P_m(\bar{x}_m));
\end{equation}
and that
$\mc{Q}$ holds in $D$ \  iff \  $D$ is inconsistent w.r.t. $\kappa(\mc{Q})$.

It holds that S-repairs are associated to causes with minimal contingency sets, while C-repairs are associated to causes for $\mc{Q}$ with minimum contingency sets, and maximum responsibilities \cite{tocs}. In fact,
for a database tuple \ $\tau \in D$:
\begin{itemize}
\item [(a)] $\tau$ \ is actual cause for $\mc{Q}$ with subset-minimal contingency set $\Gamma$ \ iff \ $D \smallsetminus (\Gamma \cup \{\tau\})$ \ is an  S-repair (w.r.t. $\kappa(\mc{Q})$),
in which case, its responsibility is \ $\frac{1}{1 + |\Gamma|}$. \item[(b)]
$\tau$ \ is actual cause with minimum-cardinality contingency set $\Gamma$ \ iff \ $D \smallsetminus (\Gamma \cup \{\tau\})$ \ is C-repair,
in which case, $\tau$ is a maximum-responsibility actual cause.
\end{itemize}
Conversely, repairs can be obtained from causes and their contingency sets \cite{tocs}. These results can be extended to unions of BCQs (UBCQs), or equivalently, to sets of denial constraints.

One can exploit the connection between causes and repairs to understand the computational complexity of the former by leveraging existing results for the latter. Beyond
the fact that computing or deciding actual causes  can be done in polynomial time  in data for CQs and UCQs \cite{suciu,tocs}, one can show that
most computational problems related to responsibility are hard, because they are also hard for repairs, in particular,  for C-repairs (all this in data complexity)
\cite{lopatenko}. In particular, one can prove \cite{tocs}: (a) The {\em responsibility problem}, about deciding if a tuple has responsibility above a certain threshold, is $\nit{N\!P}$-complete for  UCQs. \
(b) Computing  $\nit{Resp}_{_{\!D\!}}^{\!\mc{Q}}(\tau)$ \ is $\nit{F\!P}^{\nit{N\!P(log} (n))}$-complete for BCQs. This the {\em functional}, non-decision, version of the  responsibility problem. The complexity class involved is that of computational problems that use polynomial time with a logarithmic number of calls to an oracle in \nit{NP}. \ (c)
Deciding if a tuple $\tau$ is a most responsible cause is  $P^\nit{N\!P(log(n))}$-complete for BCQs. The complexity class is as the previous one, but for decision problems \cite{arora}.

\vspace{-2mm}

\section{Causes under Integrity Constraints}\label{sec:ICs}

\vspace{-2mm}
In this section we consider tuples as causes for query answering in the more general setting where  databases are subject to integrity constraints (ICs). In this scenario, and in comparison with Section \ref{sec:causal}, not every intervention on the database is admissible, because the ICs have to be  satisfied. As a consequence, the definitions of cause and responsibility have to be modified accordingly. We illustrate the issues by means of an example. More details can be found in \cite{flairsExt,kais}.

We start assuming that a database $D$ satisfies a set of ICs, $\Sigma$, i.e. $\bblue{D \models \Sigma}$. If we concentrate on BCQs, or more, generally on monotone queries, and consider causes at the tuple level, only
instances obtained from $D$ by interventions that are tuple deletions have to be considered; and they  should  satisfy the
ICs. More precisely, for $\bblue{\tau}$ \ to be actual cause for $\bblue{\mc{Q}}$, with a contingency set $\bblue{\Gamma}$, it must hold \cite{flairsExt}:
\begin{itemize}
\item[(a)] $\re{D \smallsetminus \Gamma \ \models \ \Sigma}$, \ and \ \  $\bblue{D \smallsetminus \Gamma \ \models \ \mc{Q}}$.

\item[(b)] $\re{D \smallsetminus (\Gamma \cup \{\tau\}) \ \models \ \Sigma}$, \ and \ \  $\bblue{D \smallsetminus (\Gamma \cup \{\tau\}) \ \not \models \ \mc{Q}}$.
\end{itemize}
The {\em responsibility} of $\tau$, denoted  \bblue{$\nit{Resp}_{_{\!D,\Sigma\!}}^{\!\mc{Q}}(\tau)$}, \ is defined  as in Section \ref{sec:causal}, through minimum-size contingency sets.

\begin{example} \label{ex:ics} \ Consider the database instance $D$ below, initially without additional ICs.
\begin{center}
{\footnotesize \begin{tabular}{c|c|c|} \hline
\nit{ Dep} & \nit{DName} &\nit{TStaff}  \\\hline
$t_1$& {\sf Computing} & \re{{\sf John}}   \\
$t_2$& {\sf Philosophy} &  {\sf Patrick}   \\
$t_3$&{\sf Math}  &  {\sf Kevin}   \\
 \hhline{~--} \end{tabular}~~~~ 
 \begin{tabular}{c|c|c|c|} \hline
\nit{Course}  & \nit{CName} & \nit{TStaff} & \nit{DName} \\\hline
$t_4$&{\sf COM08} & \re{\sf John}  & {\sf Computing} \\
$t_5$&{\sf Math01} & {\sf Kevin}  & {\sf Math} \\
$t_6$&{\sf HIST02}&  {\sf Patrick}   &{\sf Philosophy} \\
$t_7$&{\sf Math08}&  {\sf Eli}   &{\sf Math}  \\
$t_8$&{\sf COM01}&  \re{\sf John} &{\sf Computing} \\
 \hhline{~---}
\end{tabular} }
 \end{center}

Let us first consider the following open query:\footnote{The fact that it is open is not particularly relevant, because we can instantiate the query with the answer, obtaining a Boolean query.}

\vspace{-3mm}
\begin{equation}
\bblue{\mc{Q}(\re{x})\!: \ \exists y \exists z (\nit{Dep}(y,\re{x}) \wedge
\nit{Course(z,\re{x}, y}))}.   \label{eq:A}
\end{equation}

\vspace{-1mm}
In this case,  we get answers other that {\em yes} or {\em no}. Actually, $\bblue{\langle{\sf John}\rangle \in \mc{Q}(D)}$, the set of answers to $\mc{Q}$, and we look for causes for this particular answer. It holds: \ (a)  $\bblue{t_1}$ is a counterfactual cause; \ (b)
\gr{$t_4$ is actual cause with single minimal contingency set $\Gamma_1=\{t_8\}$}; \ (c)
$\bblue{t_8}$  is actual cause with single minimal contingency set \ $\bblue{\Gamma_2=\{t_4\}}$.

Let us now impose on $D$ the {\em inclusion dependency} (IND):

\vspace{-2mm}
\begin{equation}
\psi: \ \ \ \bblue{\forall x \forall y \ (\nit{Dep}(x, y) \rightarrow \exists u  \  \nit{Course}(u, y, x))}, \label{eq:ind}
\end{equation}

\vspace{-1mm}\noindent which is satisfied by $D$.
\ Now,
 \re{$t_4$ \ $t_8$ \ are not  actual causes  anymore}; \ and  $\bblue{t_1}$ \ is still a counterfactual cause.

Let us now consider the query: \
$\bblue{\mc{Q}_1(\re{x})\!: \ \exists y \ \nit{Dep}(y,\re{x})}$. \
Now, $\bblue{\langle{\sf John}\rangle \in \mc{Q}_1(D)}$, and
under the IND (\ref{eq:ind}), we obtain the  \re{same causes as for \ $\bblue{Q}$}, which is not surprising considering that  \  $\re{\mc{Q} \equiv_\psi \mc{Q}_1}$, i.e. the two queries are logically equivalent under (\ref{eq:ind}).

And now, consider the query:
 \
$\bblue{\mc{Q}_2(\re{x})\!: \ \exists y \exists z \nit{Course}(z,\re{x}, y)}$,
for which    $\bblue{\langle{\sf John}\rangle \in \mc{Q}_2(D)}$.
\ For this query we consider the two scenarios, with and without imposing the IND. \ Without imposing (\ref{eq:ind}), \
$\bblue{t_4}$ and $\bblue{t_8}$ are the only actual causes, with contingency sets $\bblue{\Gamma_1 = \{t_8\}}$ and $\bblue{\Gamma_2 = \{t_4\}}$, resp.

However, imposing (\ref{eq:ind}), \  $\bblue{t_4}$ and $\bblue{t_8}$ are  still  actual causes, but we lose their smallest contingency sets
$\bblue{\Gamma_1}$ and $\bblue{\Gamma_2}$ we had before:  \  $D  \smallsetminus (\Gamma_1 \cup \{ t_4\}) \not \models \psi$, \ $D  \smallsetminus (\Gamma_2 \ \cup \ \{ t_8\}) \not \models \psi$.
\  Actually, the
smallest contingency set for $\bblue{t_4}$ \ is  \ $\bblue{\Gamma_3 = \{t_8, \re{t_1}\}}$; and for
$\bblue{t_8}$, \ $\bblue{\Gamma_4 = \{t_4, \re{t_1}\}}$. \
We can see that under the IND, the responsibilities of \ $t_4$ and $t_8$ \ decrease:

\vspace{1mm}\centerline{\re{$\nit{Resp}_{_D}^{\mc{Q}_2({\sf John})}(t_4) = \frac{1}{2}$, \ and \  $\nit{Resp}_{_{D,\psi}}^{\mc{Q}_2({\sf John})}(t_4) =\frac{1}{3}$}.}

\vspace{1mm}
Tuple
 $t_1$ is not an actual cause, but it affects the responsibility of actual causes.
\boxtheorem \end{example}

Some results about causality under ICs can be obtained \cite{flairsExt}: \ (a) Causes are preserved under logical equivalence of queries under ICs, \ (b)
Without ICs, deciding causality for BCQs  is tractable, but their presence may make complexity grow. More precisely,
there are  a BCQ and an inclusion dependency
for which deciding if a tuple is an actual cause  is $\nit{N\!P}$-complete in data.

\vspace{-2mm}
\section{The Shapley Value in Databases}\label{sec:shapy}

\vspace{-2mm}
The Shapley value was proposed in game theory by Lloyd Shapley in 1953
\cite{S53}, to quantify the contribution of a player to a coalition game where players share a wealth function.\footnote{The original paper and related ones on the
  Shapley value can be found in the book edited by Alvin Roth
  \cite{R88}. Shapley and Roth shared the Nobel Prize in Economic
  Sciences 2012.} It has been applied in many disciplines. In particular, it has been investigated in computer science under
{\em algorithmic game theory} \cite{DBLP:books/cu/NRTV2007}, and it has been applied to many and
different computational problems. The
computation of the Shapley value is, in general, intractable. In many
scenarios where it is applied its computation turns out to be
$\shp$-hard \cite{FK92,DBLP:journals/mor/DengP94}. \ Here, the class $\shp$ contains the problems of {\em counting} the  solutions for problems in $\nit{NP}$. A typical problem in the class, actually, hard for the class, is $\#\nit{SAT}$, about counting the number of satisfying assignments for a propositional formula. Clearly, this problem cannot be easier than $\nit{SAT}$, because a solution for $\#\nit{SAT}$ immediately gives a solution for $\nit{SAT}$ \cite{arora}.

 Consider a set of players ${D}$,  and a
game function,  $\mc{G}:  \mc{P}(D)  \rightarrow  \mathbb{R}$, where $\mc{P}(D)$ the power set of $D$. \
 The Shapley value of player ${p}$ in ${D}$ es defined by:
  \begin{equation}{\nit{Shapley}(D,\mc{G},p):= \sum_{S\subseteq
  D \setminus \{p\}} \frac{|S|! (|D|-|S|-1)!}{|D|!}
(\mc{G}(S\cup \{p\})-\mc{G}(S))}.\label{eq:sh}\end{equation}
  Notice that here, ${|S|! (|D|-|S|-1)!}$ is the number of permutations of
${D}$ with all players  in ${S}$  coming first, then ${p}$, and then all the others. That is, this quantity
is the expected contribution of player $p$ under all possible additions of $p$ to a partial random sequence of players followed   by a random sequence of the rests of   the players. Notice the counterfactual flavor, in that there is a comparison between what happens having $p$ vs. not having it. The Shapley value is the only function that satisfies certain natural properties in relation to games. So, it is a
result of a categorical set of axioms or conditions \cite{R88}.

The Shapley value has been used in knowledge
representation, to measure the degree of inconsistency of a
propositional knowledge base \cite{HK10}; in machine learning to provide explanations for the outcomes of
classification models on the basis of numerical scores assigned to the participating feature values \cite{LL17,LetA20}. It has also been applied in data management to
measure the contribution of a tuple to a query answer \cite{LBKS20,SigRec21}, which we briefly review in this section.

In databases, \ the players are tuples in a database  ${D}$. We also have a
Boolean query \ $\mc{Q}$, which becomes a   game function, as follows:  \ For \ ${S \subseteq D}$, i.e. a subinstance,

\vspace{-2mm}
\begin{equation*}{\mc{Q}(S) = \left\{\begin{array}{cc} 1 & \mbox{ if } \ S \models \mc{Q},\\
0 & \mbox{ if } \ S \not \models \mc{Q}.\end{array}\right.}
\end{equation*}

\vspace{-1mm}
With these elements we can define the Shapley value of a tuple $\tau \in D$:

\vspace{-3mm}
\begin{equation*}
\nit{Shapley}(D,{\mc{Q}},{\tau}):= \sum_{S\subseteq
  D \setminus \{{\tau}\}} \frac{|S|! (|D|-|S|-1)!}{|D|!}
(\mc{Q}(S\cup \{{\tau}\})-\mc{Q}(S)).
\end{equation*}

\vspace{-2mm}
If the query is {\em monotone}, i.e. its set of answers never shrinks when new tuples are added to the database, which is the case of conjunctive queries (CQs), among others, the difference $\mc{Q}(S\cup \{{\tau}\})-\mc{Q}(S)$ is always $1$ or $0$, and the average in the definition
of the Shapley value returns a value between $0$ and $1$. \
This value quantifies the contribution of tuple \ ${\tau}$ \ to the query result. It was introduced and investigated in \cite{LBKS20,SigRec21},
for BCQs and some aggregate queries defined over CQs. We report on some of the findings in the rest of this section. The analysis has been extended to queries with  negated atoms in CQs \cite{ester2}.

A main result obtained in \cite{LBKS20,SigRec21} is about the complexity of computing this Shapley score. The following
{\em Dichotomy Theorem} holds:  \ For ${\mc{Q}}$ a BCQ without self-joins, if \ ${\mc{Q}}$ \ is {\em hierarchical}, then ${\nit{Shapley}(D,\mc{Q},\tau)}$ can be computed in polynomial-time
(in the size of $D$); otherwise, the problem is \ {$\nit{\#P}$-complete}.

Here,  ${\mc{Q}}$ \ is {hierarchical} if for every two existential variables ${x}$ and ${y}$, it holds: \ (a)
{$\nit{Atoms}(x) \subseteq \nit{Atoms}(y)$}, \ or
 {$\nit{Atoms}(y) \subseteq \nit{Atoms}(x)$}, \ or
 {$\nit{Atoms}(x) \cap \nit{Atoms}(y) = \emptyset$}.
 \ For example,  ${\mc{Q}: \ \exists x \exists y \exists z(R(x,y) \wedge S(x,z))}$, for which \
{$\nit{Atoms}(x)$ $ = \{R(x,y),$ $ \ S(x,z)\}$,  \ $\nit{Atoms}(y) = \{R(x,y)\}$,  \ $\nit{Atoms}(z) = \{S(x,z)\}$}, is
hierarchical. \
However, \ ${\mc{Q}^{\nit{nh}}: \ \exists x \exists y({R(x) \wedge S(x,y) \wedge T(y)})}$, for which \
  {$\nit{Atoms}(x) = \{R(x), \ S(x,y)\}$,  \ $\nit{Atoms}(y) = \{S(x,y), T(y)\}$}, is not hierarchical.

These are the same criteria for (in)tractability that apply to evaluation of BCQs  over probabilistic databases \cite{probDBs}. However, the same proofs do not  apply, at least not straightforwardly.
The intractability result uses query  ${\mc{Q}^{\nit{nh}}}$ \ above, and a
reduction from {counting independent sets in a bipartite graph}.

The {dichotomy results can be extended  to summation} over CQs, with the same conditions and cases. This is because the
Shapley value, as an expectation, is linear. \
{Hardness extends to aggregates {\sf max}, {\sf min}, and {\sf avg} over non-hierarchical queries}.

For the hard cases, there is, as established in \cite{LBKS20,SigRec21},  an {\em approximation result}: \ For every fixed BCQ $\mc{Q}$ (or summation over a  CQ), there is a {\em multiplicative fully-polynomial randomized approximation scheme} (FPRAS) \cite{arora}, $A$, with, for given $\epsilon$ and $\delta$:

\vspace{-4mm}
$${P(\tau \in D ~|~ \frac{\nit{Shapley}(D,\mc{Q},\tau)}{1+\epsilon} \leq A(\tau,\epsilon,\delta) \leq (1 + \epsilon)\nit{Shapley}(D,\mc{Q},\tau)\}) \geq 1 - \delta}.$$

\vspace{-2mm}
 A related and popular score, in coalition games and other areas, is the  {\em Bahnzhaf Power Index}, which is similar to the Shapley value, but the order of players is ignored, by considering subsets of players rather than permutations thereof. It is defined by:

 \vspace{-2mm}
\begin{equation*}{\nit{Banzhaf}(D,\mc{Q},\tau) := \frac{1}{2^{|D|-1}} \cdot \sum_{S \subseteq (D\setminus \{\tau\})} (\mc{Q}(S \cup \{\tau\}) - \mc{Q}(S))}.
\end{equation*}

\vspace{-1mm}
The Bahnzhaf-index is  also difficult to compute; provably \#{\it P}-hard in general. The results in \cite{LBKS20,SigRec21} carry over to this index when applied to query answering.
\ In \cite{LBKS20} it was proved that the causal-effect score of Section \ref{sec:CE} coincides with the Banzhaf-index, which gives to the former an additional justification.

\vspace{-2mm}

\section{Final Remarks}\label{sec:last}

\vspace{-2mm}
 Explainable data management and explainable AI  (XAI) are effervescent areas of research.
 The relevance of explanations can only grow, as observed from- and due to the legislation and regulations that are being produced and enforced in relation to explainability, transparency and fairness of data management and AI/ML systems.

There are different approaches and methodologies in relation to explanations, with causality, counterfactuals and scores being prominent approaches that have a relevant role to play.
Much research is still needed on the use of {\em contextual, semantic and domain knowledge}. Some approaches may be more appropriate in this direction, and we argue that declarative, logic-based specifications can be successfully exploited \cite{tplp}.

 Still fundamental research is needed in relation to the notions of {\em explanation} and {\em interpretation}. An always present question is: {\em What is a good explanation?}. \ This is not a new question, and in AI (and other areas and disciplines) it has been investigated. In particular in AI,  areas such as {\em diagnosis} and  {\em causality} have much to contribute.

Now, in relation to {\em explanations scores}, there is still a question to be answered: \ {\em What are the desired properties of an explanation score?}. The question makes a lot of sense, and may not be beyond an answer. After all, the  general
Shapley value emerged from a list of {\em desiderata} in relation to coalition games, as the only measure that satisfies certain explicit properties \cite{S53,R88}. Although the Shapley value is being used in XAI, in particular in its $\nit{Shap}$  incarnation, there could be a different and specific  set of desired properties of explanation scores that could lead to a still undiscovered explanation score.

 \vspace{3mm} \noindent {\bf Acknowledgments: }  \ Part of this work was funded by ANID - Millennium Science Initiative Program - Code ICN17002; and NSERC-DG 2023-04650.

\end{document}